# Bell and von Neumann

Jeremy Bernstein

Stevens Institute of Technology

Abstract: This article discusses von Neumann's "proof" that hidden variable quantum theories are impossible.

I spent the spring of 1989 at CERN in Geneva. I had two activities one was overt and the other covert in a manner of speaking. The overt activity was to deliver a set of lectures to a general lab audience on neutrino cosmology, a subject that was becoming fashionable. The covert activity was to do a series of Interviews with John Bell that I hoped would become a New Yorker magazine profile. I was then on the staff of the New Yorker and over the years had done profiles of people like Hans Bethe, I.I,Rabi and Einstein. But we had a fairly new editor Robert Gottlieb and I was not sure how much interest he had in scientific profiles of the kind that I did. Bell was at that time well-known among physicists but he had not yet become the cult figure that he became;even more so after his death in 1990. I had a tape recorder and used it for a series of interviews that went on for several hours. I have now turned the tapes over to CERN and an attempt, only partially successful, has been made to improve their quality. They were good enough so that I could use them to write my profile. I was not entirely surprised when Gottlieb turned it down. He had no sense of the importance of the physics-perhaps I failed to convey that-and he had no sense of Bell as a person. Bell died of a cerebral aneurysm in the fall of 1990. That year he had been nominated for a Nobel Prize which I think he would have won. In the event I put my profile in a book called Quantum Profiles[1] pretty much as I wrote it for the New Yorker. But there was something I skipped over lightly largely because I had not taken the trouble to try to understand it. That is von Neumann's famous "proof" that a deterministic theory that reproduces the results of quantum mechanics is impossible. Showing the limitations of this argument became a cornerstone of Bell's understanding of the quantum theory. I want to make up for this oversight in this essay which I dedicate to his memory.

---

[1] Princeton University Press, Princeton, 1991

Bell began reading popular books on the quantum theory when he was still in high school. In 1945 he entered Queen's University in Belfast and took his first formal course in the quantum theory. The instructor was one Robert Sloane. On feels some sympathy for him. In my experience most students who begin quantum theory are in a state of shock and awe. That was certainly how I felt when took my first course with Julian Schwinger. This was enhanced by the fact that Schwinger was trying out on us his then new Green's function approach to the theory. We did not see the Schrödinger equation until some time in the second semester. Many of us went to MIT to audit Vicky Weisskopf's more earth-bound lectures. I cannot imagine what either of them would have done with the young John Bell. At the time he was particularly irritated by the uncertainty principles, As he recalled, "It looked as if you could take this size and then the position is well-defined or that size and the momentum is well-defined. It sounded as if you were just free to make it what you wished. It was only slowly that I realized that it is not a question of what you wished, It is really a question of which apparatus has produced this situation. But for me it was a bit of a fight to get through to that, It was not very clearly set out in the books and courses that were available to me. I remember arguing with one of my professors, a Doctor Sloane about that. I was getting very heated and accusing him, more or less ,of dishonesty. He was getting heated too and said 'You're going too far.' But I was very engaged and angry that we couldn't get all that clear."[2] Poor Sloane-he must have felt that he was being attacked by a red-haired dervish. I asked Bell if at the time he thought the theory might actually be wrong, "I hesitated to think it might be wrong" he said, "but I *knew* it was rotten.That is to say, one has to find some decent way of expressing whatever truth there is in it."[3] I wish that I could provide the CD with this essay so that you could hear the relish with which Bell pronounces the word "rotten."

---

[2] Bernstein op cit pp 50-51.

It rapidly became clear to Bell that the notion of 'apparatus' was central to the usual interpretation of the quantum theory. Moreover Bohr had insisted that anything that was an apparatus had to be describable in terms of classical physics. If not, you seemed to get into an infinite regress. But this raised the question of where the division was between the classical and quantum worlds. Bohr never made this clear and neither did any of his followers although as a practical matter this did not seem to pose a problem. I remember in an interview I did with I.I. Rabi he said that Bohr was very profound about things that did not matter, They mattered very much to Bell. Enter Max Born. Born was certainly one of the most distinguished theoretical physicists of the 20[th] century. Among many things, we owe to him the probabilistic interpretation of the Schrödinger wave function. For this he was awarded the Nobel Prize in physics in 1954, pretty late for work that was done in the 1920s, In 1933 he was forced to leave Germany and he settled in Great Britain-at the University of Edinburgh finally. But in 1948 he delivered the Waynflete Lectures at the St,Mary Magdelin college in Oxford. They were entitled "Natural Philosophy of Cause and Chance" and published under that title the following year.[4] He naturally discussed the quantum theory and it was the following that got Bell's attention.

"It would be silly and arrogant to deny any possibility of a return to determinism. For no physical theory is that final; new experiences may force us to alternatives and even revisions. Yet scanning the history of physics in the way we have done we see fluctuations and vacillations, but hardly a reversion to more primitive conceptions. I expect that our present theory will be profoundly modified. For it is full of difficulties which I have not mentioned at all- the self-energies of particles in interaction and many other quantities, like collision cross cross-sections, lead to divergent integrals. But I should never expect that these difficulties could be solved by a return to classical concepts. I expect just the opposite, that we shall have to sacrifice some current ideas and use still more abstract methods."

---

[3] Bernsetin,op cit, 20
[4] Oxford University Press. Oxford, 1949. But see the revised 1964 Dover Press Edition, New York, I take page numbers from this edition.

Then came the passage that really got Bell's attention.

"However, these are only opinions. A more concrete contribution to this question has been made by J.v.Neumann in his brilliant book ,*Matematische Grundlagen der Quantenmechanik*. He puts the theory on an axiomatic basis by deriving it from a few postulates of a very plausible and general character about the properties of 'expectation values' (averages) and their representation by mathematical symbols. The result is that the formalism of quantum mechanics is uniquely determined by these axioms; in particular no concealed parameters can be introduced with the help of which the indeterministic description could be transformed into a deterministic one. Hence if a future theory should be deterministic, it cannot be a modification of the present one, but must be essentially different. How this should be possible without sacrificing a whole treasure of well-established results I leave to the determinists to worry about."[5]

This seemed to settle the matter.

Bell graduated from Queen's University in 1949 and immediately took a job designing accelerators for the Atomic Energy Research Establishment in Malvern. Von Neumann's book had not yet been translated from German and besides Bell was occupied by his work although he still kept thinking about quantum mechanics. But early in 1952 something happened that changed everything. David Bohm published two papers which did just what von Neumann had said was impossible. He created a hidden variable theory which was at least in some part deterministic and which reproduced all the results of the quantum theory. Something must have been terribly wrong with von Neumann. There was a physicist of German origin at Malvern named Franz Mandl. He could of course read German and he had an interest in the foundations of the quantum theory. He was able to tell Bell what von Neumann had done and from this Bell got an idea of where the problem was. If you look at von Neumann's book-I have the

---

[5] Born, op cit, pp 108-109

English translation Mathematical Foundations of Quantum Mechanics[6] you will be able with some scrutiny to look at what seems to be the relevant discussion on page 324. I put things this way because my guess is that you will be baffled. Basil Hiley told me that he and Bohm studied this for a few years and never completely understood what von Neumann was doing until Bell explained it. This is what I propose to do. Sometimes people write that von Neumann made a "mistake." Von Neumann did not make mistakes. What he did is perfectly correct but it is totally irrelevant.

The expectation value of an operator A in a state ψ is given by

$$\langle A \rangle = \int \Psi^* A \Psi dx.$$

and the dispersion is defined to be

$$(\Delta A)^2 = \int \Psi^* (A - \langle A \rangle)^2 \Psi dx,$$

Since the integrand is positive definite, for the dispersion to vanish, ψ must be an eigen-state of A with eigen-value $\langle A \rangle$. The second thing to note is that is if A and B are any two possibly non-commuting operators then

$$\langle A + B \rangle = \langle A \rangle + \langle B \rangle.$$

Von Neumann insists that this averaging property must be a characteristic of any proposed hidden variable theory. But if this theory is to be deterministic then he argues there must be at least one state that is dispersion free for all observables. This state must be an eigen-state of A and B as well as A+B. By the additivity assumption it must be that for all such operators the sum of the eigen-values of the two operators is the eigen-value of the sum. But this statement is generally false. A simple example is supplied by the Pauli σ matrices. The eigen- values of $\sigma_x$ and $\sigma_y$ are ±1 while the eigen- values of $\sigma_x + \sigma_y$ are $\pm\sqrt{2}$. Thus von Neumann has made his point but it has nothing to do with Bohmian mechanics.

---

[6] Princeton University Press, Princeton, New Jersey, 1955

In Bohmian mechanics the wave function ψ obeys a Schrödinger equation. But it is this wave function that is, if you like, the hidden variable. There are particles that move along deterministic trajectories where the trajectories are determined by a quantum mechanical version of Newton's law. For a single particle with a position Q this equation of motion reads

$$m\ddot{Q} = -\nabla(V(Q) + U(Q)).$$

Here V is the classical potential and U is the quantum potential which is given by

$$U = -\hbar^2/2m(\nabla^2 R/R),$$

where $R = \sqrt{\Psi^*\Psi}$. As is evident, von Neumann's critique has nothing to do with the case. Once Bell realized this he was free to go on and explore the theory. He noted that Bohmian mechanics is non-local. If for example there are two particles in interaction then the values of the potentials for one of them depend on the variables of the second at the same time no matter how distant the two points are from each other. This is an example of what Einstein called in a letter to Born *spukhafte Fernwirkungen*- spooky actions as a distance. This did not bother Bell in the least. This kind of non-locality is an ineluctable aspect of the quantum theory and Bohmian mechanics makes it explicit. What concerned Bell was relativity. Bohmian mechanics is non-relativistic. Indeed the marriage of relativity and the quantum theory is an uneasy one. In the usual version, measurements involve the instantaneous collapse of the wave function which is not a Lorentz covariant concept. Bell followed the work on the interpretation of the quantum theory closely. He was not happy with any of them although Bohmian mechanics came the closest.

Jeremy Bernstein